 \documentclass[final,5p,times,twocolumn]{elsarticle}

\usepackage{amssymb}
\usepackage{amsmath}
\usepackage{wasysym}
\usepackage{aas_macros}
\usepackage{natbib}
\usepackage{url}
\newcommand{\hr}{\texttt{HR}}

\newcommand{\ms}{\texttt{MS}}
\newcommand{\tof}{\texttt{TO}}
\newcommand{\oc}{\texttt{OC}}
\newcommand{\ad}{\texttt{AD11}}
\newcommand{\nacre}{\texttt{NACRE99}}
\newcommand{\msv}{\texttt{MSV13}}
\newcommand{\jina}{\texttt{JINA}}
\newcommand{\prosecco}{\texttt{PROSECCO}}
\newcommand{\ssm}{\texttt{SSM}}
\newcommand{\second}{\prime\prime}

\journal{Physics Letters B}

\begin{document}

\begin{frontmatter}



\title{Astrophysical implications of the proton-proton cross section updates}
\author[a,c]{E. Tognelli}
\ead{tognelli@df.unipi.it}
\author[b,c]{S. Degl'Innocenti}
\ead{scilla@df.unipi.it}
\author[b,c]{L. E. Marcucci}
\author[b,c]{P.G. Prada Moroni}
\address[a]{Department of Physics, University of Roma Tor Vergata, Via della Ricerca Scientifica 1, 00133, Roma, Italy}
\address[b]{Department of Physics 'E.Fermi', University of Pisa, Largo Bruno Pontecorvo 3, 56127, Pisa, Italy}
\address[c]{INFN, Section of Pisa, Largo Bruno Pontecorvo 3, 56127, Pisa, Italy}



\begin{abstract}
The p(p,e$^+ \nu_e)^2$H reaction rate is an essential ingredient for theoretical computations of stellar models. In the past several values of the corresponding $S$-factor have been made available by different authors. Prompted by a recent evaluation of $S(E)$, we analysed the effect of the adoption of different proton-proton reaction rates on stellar models, focusing, in particular, on the age of mid and old stellar clusters (1-12 Gyr) and on standard solar model predictions. By comparing different widely adopted p(p,e$^+ \nu_e)^2$H reaction rates, we found a maximum difference in the temperature regimes typical of main sequence hydrogen-burning stars ($5\times 10^6 \div 3\times 10^7$ K) of about $3\%$. Such a variation translates into a change of cluster age determination lower than $1\%$. A slightly larger effect is observed in the predicted solar neutrino fluxes with a maximum difference, in the worst case, of about 8$\%$. Finally we also notice that the uncertainty evaluation of the present proton-proton rate is at the level of few $\permil$, thus the p(p,e$^+ \nu_e)^2$H reaction rate does not constitute anymore a significant uncertainty source in stellar models.
\end{abstract}


\end{frontmatter}


\section{Introduction}
\label{sec:introduction}

At the energies of interest for stellar nucleosynthesis the rate of the proton-proton (p-p) weak capture is too low to be directly measured in laboratory and it can be determined only by means of nuclear physics calculations. The p(p,e$^+ \nu_e$)$^2$H reaction drives the efficiency of the p-p chain, which is fundamental for hydrogen burning in stars. Thus, a variation of the p-p reaction rate adopted in the stellar models potentially influences the characteristics and the evolutionary times (at least) during the central hydrogen burning phase (Main Sequence, \ms) of low-mass stars and thus the age determination of old stellar clusters (see e.g. discussions in Refs.~\cite{chab96,brocato98,valle13a,valle13b}, and references therein). Moreover, as the Sun burns hydrogen mainly by means of the p-p chain, a change in the p-p cross section adopted in the models can affect the predicted solar structure and, consequently, both the neutrino fluxes and the helioseismological observables.

The p-p reaction rate is expressed in terms of the astrophysical $S$-factor $S(E)$ \cite{fowler67,adelberger11}, where the energy, $E$, is measured in the two-proton centre-of-mass frame.  $S(E)$ is often expressed as the first three terms of a Maclaurin series in $E$:

\begin{equation}
S(E)=S(0)+S^\prime(0) \times E + \frac{1}{2}S^{\second} (0) \times E^2 + \dots
\label{eq:SE}
\end{equation}
where $S^\prime(0)$ and $S^{\second}(0)$ are the first and second derivatives of $S(E)$ evaluated at zero energy. The recent reaction rate compilation by Adelberger et al. \cite{adelberger11} reviewed in a detailed way different evaluations for $S(0)$ and $S^\prime(0)$ (see also the discussion in Ref.~\cite{marcucci13}), recommending the values: S(0)=(4.01 $\pm$ 0.04) $\times 10^{-25}$ MeV\,b and $S^\prime(0)/S(0)=(11.2 \pm 0.1)$ MeV$^{-1}$. The \nacre{} \cite{angulo99} compilation, still widely adopted in the literature, suggested at the time: $S(0)$=3.94 $\times 10^{-25}$ MeV\,b, $S^\prime(0)$/$S(0)$=11.7 MeV$^{-1}$ and $S^{\second}(0)$/$S(0)$=(150 $\pm$ 20) MeV$^{-2}$.

Recently, the astrophysical $S$-factor has been calculated by Marcucci et al. \cite{marcucci13} (hereafter \msv) applying the so-called chiral effective field theory framework, which allows for a better determination of the theoretical uncertainty. Taking into account two-photon and vacuum polarisation contributions beyond simple Coulomb interaction, and, moreover, the contributions of the S- and P-partial waves in the initial p-p state, the corresponding value for $S(0)$ has been found to be (4.030 $\pm$ 0.006) $\times 10^{-25}$ MeV\,b, with the uncertainty reduced by about a factor seven with respect to previous evaluations. Note that the P-partial waves have been found to give a contribution of about 0.02 $\times 10^{-25}$ MeV b, explaining the difference with the $S(0)$ value of Ref.~\cite{adelberger11}. Marcucci et al. \cite{marcucci13} provided values for $S^\prime(0)$, $S^{\prime\prime}(0)$ and even higher derivatives to be used in Eq. (\ref{eq:SE}), by fitting $S(E)$ in the 0-100 keV energy range. However, we preferred to directly obtain the p-p rate in the required energy range and with the chosen energy resolution by using a routine, based on the $S(E)$ evaluated in Ref.~\cite{marcucci13}, which is made available at the link  \url{http://astro.df.unipi.it/stellar-models/pprate/}.

In this work, relevant stellar evolutionary quantities calculated by adopting the Marcucci et al.~\cite{marcucci13} p-p reaction rate with the inclusion of the S- and P-partial waves (hereafter \msv\texttt{(S+P)}) are compared with those obtained by using different p-p rates widely adopted in the literature, and with the \msv{} rate calculated without the inclusion of the P-partial wave contribution (hereafter \msv\texttt{(S)}). We concentrate on the age determination of stellar clusters and on the solar model characteristics, including the solar neutrino fluxes.

Section~\ref{sec:rates} describes the calculation of the p(p,e$^+\nu_e$)$^2$H reaction rate, with the related uncertainty; the obtained reaction rate is then compared with previous evaluations. In Sec.~\ref{sec:models} the characteristics of the stellar models and of the adopted evolutionary code are briefly described, while in Sec.~\ref{sec:isochrones} and Sec.~\ref{sec:sun} the effects of the p-p rate update on age estimation of stellar clusters and on standard solar model characteristics are discussed. We conclude with a summary in Sec.~\ref{sec:conclusion}.

\section{The  p(p,e$^+ \nu_e)^2$H reaction rate}
\label{sec:rates}

The rate $R$ for the p-p reaction is commonly expressed in cm$^3$ mol$^{-1}$ s$^{-1}$, such that (see e.g. Ref.~\cite{NACREII})
\begin{eqnarray}
R&\equiv& N_A\langle\sigma v\rangle = \frac{3.73\times 10^{10}}{\sqrt{\hat{\mu}\,T_9^{3}}} \times \nonumber\\
&&\times \int_0^\infty S(E)\,{\rm exp}\bigg(-2\pi\eta -11.605 \frac{E}{T_9}\bigg)\,\mathrm{d}E
\label{eq:rate}
\end{eqnarray}
where $N_A$ is the Avogadro number, $ \langle\sigma v\rangle$ is the Maxwellian-average rate, ${\hat{\mu}}=0.504$ is the p-p reduced mass in atomic mass units (1 amu = 931.494 MeV/c$^2$), $T_9$ is the temperature in units of $10^9$ K, and $\eta$ is the Sommerfeld parameter expressed as $\eta=0.1575(\frac{\hat{\mu}}{E})^{1/2}$. The integration over the centre-of-mass energy $E$ in Eq.~(\ref{eq:rate}) can be performed numerically with standard techniques. The crucial input in this calculation is $S(E)$. In Ref.~\cite{marcucci13}, $S(E)$ has been calculated in the range $E=0$-100 keV, with the inclusion of S- and P-partial waves in the initial p-p state. A realistic estimation of the theoretical uncertainty can be performed, since $S(E)$ is calculated within the so-called chiral effective field theory framework \cite{marcucci13}, which allows to reduce the theoretical uncertainty to the order of few $\permil$, by constraining systematically both the nuclear potential and the nuclear current operator with a stringent contemporary fit of the trinucleon binding energy and tritium $\beta$-decay lifetime. This is better than what has been done in Ref.~\cite{adelberger11}, where the study in the ``old-fashion'' 
potential model approach of Ref.~\cite{schiavilla98} was primary used, and the uncertainties most of all arising from two- and three-body potentials and currents lead to an accuracy for $S(0)$ not better than 1 \%.
Using the uncertainty on $S(E)$ of Ref.~\cite{marcucci13}, we obtain an upper and lower value for $S(E)$ at a given value of $E$.
Consequently it is possible to derive an upper and lower value for the rate $R$, or, alternatively, to provide a theoretical uncertainty $\Delta R$. Two more options are present in the on-line release of the computer program which calculates $R$: (i) the contributions from the P-partial waves of the initial p-p state can be excluded; (ii) rather than using the calculated values of $S(E)$, tabulated on 101 grid points with steps of 1 keV starting from $E=0$, it is possible to use the Maclaurin series in $E$ (see Eq.~(\ref{eq:SE})), with $S(0)$, $S^\prime(0)$, $S^{\prime\prime}(0)$ and $S^{\prime\prime\prime}(0)$ given by Marcucci et al.~\cite{marcucci13}, and fitted to reproduce $S(E)$ in the 0-100 keV range (see Tab.\ref{tab:values}). Then, the energy range and energy resolution can be chosen as preferred. No appreciable difference is seen in the calculation of $R$ by using the 101 grid point of $S(E)$ or the fitted values of $S(0)$ and its derivatives. For a better comparison of the differences among the selected evaluations of the p-p rate, and for future reference, we report in Table~\ref{tab:values} the values of $S(0)$, $S^\prime(0)$, $S^{\prime\prime}(0)$ and $S^{\prime\prime\prime}(0)$ as obtained by Marcucci et al.~\cite{marcucci13}, compared with those available in the literature, namely in the \nacre{} \cite{angulo99} and \ad{} \cite{adelberger11} compilations. It is to be noticed that, as already mentioned in Ref.~\cite{marcucci13}, a good fit of $S(E)$ with the Maclaurin series in $E$ in the range 0-100 keV is obtained only including up to the third derivative $S^{\prime\prime\prime}(0)$.

\begin{figure}[!t]
   \centering
   \includegraphics[width=0.9\columnwidth]{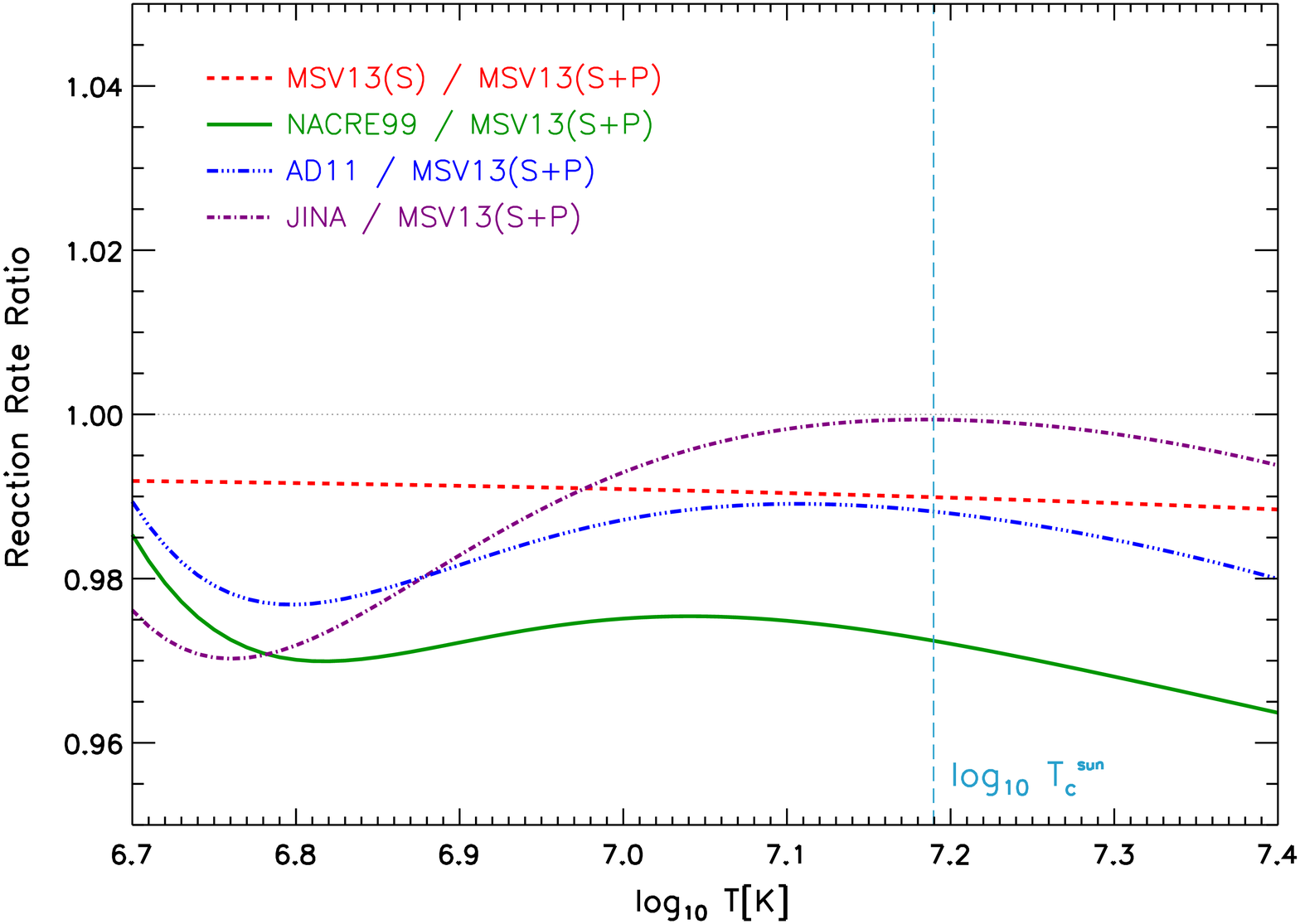}
   \caption{Proton-proton reaction rate ratios between different compilations present in the literature and our reference one, \msv\texttt{(S+P)}, based on the Marcucci et al. \cite{marcucci13} $S(E)$ calculation. \msv\texttt{(S)} indicates the rate calculated by adopting the Marcucci et al. $S(E)$ evaluation without the inclusion of P-partial wave contributions. The vertical dashed line marks the solar central temperature.}
 \label{rates}%
\end{figure}

Figure \ref{rates} shows the ratio between the present p(p,e$^+\nu_e$)$^2$H reaction rate calculated by adopting the $S(E)$ evaluation of Ref.~\cite{marcucci13} and those reported by the \nacre{} \cite{angulo99}, \ad{} \cite{adelberger11}, and \jina{} \citep{cyburt10}, widely used in the literature. The ratio between our reference choice (\msv\texttt{(S+P)}) and the p-p reaction rate calculated without the inclusion of the P-partial wave contribution (\msv\texttt{(S)}) is also shown. The results are shown only in the temperature range of astrophysical interest for central hydrogen-burning stars.

\begin{table*}[!ht]
	\centering
	\begin{tabular}{l|cccc}
                            & $S(0)$ & 
                              $S^\prime(0)/S(0)$  & 
                              $S^{\prime\prime}(0)/S(0)$ &
                              $S^{\prime\prime\prime}(0)/S(0)$ \\
                            & [$\times 10^{-23}$ MeV fm$^2$] & 
                              [MeV$^{-1}$]        & 
                              [MeV$^{-2}$] &
                              [MeV$^{-3}$] \\
        \hline
        \hline
\msv\texttt{(S+P)} & 4.030(6) & 11.94(1) & 248.8(2) & -1183(8) \\
\msv\texttt{(S)}   & 4.008(5) & 11.42(1) & 239.6(5) & -1464(5) \\ 
\nacre             & 3.94     & 11.7     & 150(20)  &          \\
\ad                & 4.01(4)  & 11.2(1)  &          &          \\
	\hline
	\hline
	\end{tabular}
	\caption{Values for $S(0)$ and its first, second and third derivatives as obtained by Marcucci et al.~\cite{marcucci13}, including (\msv\texttt{(S+P)}) and not including (\msv\texttt{(S)}) the P-partial wave contribution. The corresponding values for the reaction rate compilations quoted in Sec.~2 are also listed (if available in the literature).}
\label{tab:values}
\end{table*}

In the temperature range of interest, the largest relative variation of the reaction rate (about 3-4$\%$) with respect to the reference one is found for the \nacre{} compilation, which adopts an $S(0)$ value different by about $2\%$ from the \msv\texttt{(S+P)}. The temperature behaviour of the \jina{} and \ad{} rate is different from the present one, but the relative changes remain within $\approx 2\%$ (\ad) or $\approx 2$-3$\%$ (\jina) for the whole selected temperature range. The effect on the \msv{} rates of the inclusion of the P-partial wave contribution is of the order of $1\%$. 
It is also worth to discuss the differences among the various rates in the light of the quoted uncertainties. Notice that the error of the present \msv{} p-p reaction rate is of the order of few $\permil$. 
The difference between the \nacre{} and the present reaction rate in the temperature range of interest is smaller than the \nacre{} estimated uncertainty, which ranges between $-3$\% and $+7$\% (see Table 2 in Ref. \cite{angulo99}). However, it has to be emphasised that the values of $S(0)$ and its derivatives adopted by \nacre{} appear to be outdated in the light of more recent calculations; thus, the adoption of the \nacre{} p-p rate should be avoided. The differences between \ad{} and  \msv{} reaction rates are larger than the uncertainty on $S(0)$ quoted by \ad{} (about $1\%$). However, the global uncertainty on \ad{} $S(E)$ also depends on the error on $S^\prime(0)$ and on the lack of the second derivative $S^{\prime\prime}(0)$. Numerical simulations show that the lack of $S^{\prime\prime}(0)$ affects the resulting reaction rate by less than $1\%$, in the temperature range of interest (see also Ref. \cite{bahcall69}). Finally we are not able to discuss the differences between the \jina{} and \msv{} rates because no information about the uncertainty on the \jina{} rate are available on the \jina{} web page.

\section{Stellar models}
\label{sec:models}

Stellar models have been calculated by means of the \prosecco{} stellar evolutionary code \citep{deglinnocenti08,tognelli14}. A detailed discussion of the adopted input physics can be found in Refs.~\cite{tognelli11,tognelli12,dellomodarme12}; here we just summarise the most recent updates. 

The main novelty concerns the adopted nuclear reaction rates relevant for the present work. In particular the recent \ad{} rates have been adopted in place of the \nacre{} ones for the following reactions: $^3$He($^3$He,2p)$^4$He, $^3$He($^4$He,$\gamma$)$^7$Be, p($^7$Be,$^4$He)$^4$He, p($^{12}$C,$\gamma$)$^{13}$N, and p($^{16}$O,$\gamma$)$^{17}$F. For the bottleneck reaction of the CNO cycle p($^{14}$N,$\gamma$)$^{15}$O, we adopted the \texttt{LUNA} results \cite{imbriani05}.

Bare nuclei reactions have been corrected to account for the plasma electron screening for weak \cite{salpeter54}, weak-intermediate-strong \cite{dewitt73,graboske73}, and strong \cite{itoh77,itoh79} screening. Only in the case of standard solar model calculations (see Sec.~\ref{sec:sun}), following the choice of most authors, the Salpeter formula for weak-screening is adopted for all the p-p chain and CNO-cycle reactions (see e.g. Ref.~\cite{bahcall02}). Atomic diffusion has been included, taking into account the effects of gravitational settling and thermal diffusion, with coefficients given by Thoul et al. \cite{thoul94}.

Regarding the chemical composition, models have been calculated for two initial [Fe/H]\footnote{$[\mathrm{Fe/H}] \stackrel{def}{=} \log \frac{(N_{\mathrm{Fe}}/N_{\mathrm{H}})_\star}{(N_{\mathrm{Fe}}/N_{\mathrm{H}})_\odot}$.} values, namely [Fe/H]=$+0.0$ (corresponding to the solar composition) and [Fe/H]=$-1.0$, to take into account the range of values for this quantity observed in Galactic stars. [Fe/H] has been converted into initial helium ($Y$) and metal ($Z$) mass fractions by adopting eqs.~(1) and (2) in Ref.~\cite{gennaro10}, which is valid for solar-scaled metal distribution. Present models have been computed by adopting the recent solar metals abundances given by Asplund et al. (2009) \cite{asplund09}. The initial $Y$ and $Z$ for the two sets of [Fe/H] values are: ($Y$, $Z$)=(0.274, 0.013) for [Fe/H]=$+0.0$ and ($Y$, $Z$)=(0.250, 0.001) for [Fe/H]=$-1.0$.

It is worth to emphasise that all the analysis presented in this paper have been performed in a differential way, i.e., the results obtained with the reference p(p,e$^+ \nu_e$)$^2$H reaction rate (\msv\texttt{(S+P)}) have been compared with those obtained with the other evaluations of the p-p rate discussed above, keeping all the other physical parameters and the stellar chemical composition fixed. Thus, the results are expected to be weakly dependent on the chemical composition and on the input physics adopted in the models (see Refs.~\cite{valle13a,valle13b}).

\section{Proton-proton cross section and age determination in stellar clusters}
\label{sec:isochrones}

Stellar clusters provide a severe benchmark for stellar evolution models, since they consist of stars sharing the same age, chemical composition and distance from the Earth. The age determination of stellar clusters in the Milky Way and in the other galaxies is of paramount importance, as it provides information about the evolutionary history of the host galaxies. The age of the stellar clusters is determined through the luminosity of the point corresponding to the central H-exhaustion (Turn-Off, \tof, for old clusters and Overall Contraction, \oc, for the younger ones). The older is the cluster and the lower is the mass at the \tof /\oc{} point, and, thus, the lower is its luminosity. Therefore, the efficiency of the p-p reaction is relevant for stellar cluster dating. For stars in clusters older than about 10 Gyr, the p-p chain is the main H-burning channel during the \ms{} phase, while for stars close to the \tof{} phase in younger clusters the CNO-cycle dominates, although the p-p chain contribution remains not negligible at least for ages older than about 1 Gyr. This roughly corresponds to masses of about 1.5 M$_{\odot}$ (for solar chemical composition). We will not discuss post-\ms{} phases as the contribution of the p-p chain becomes negligible with respect to the CNO-cycle one.

We computed stellar evolutionary tracks\footnote{The evolutionary track is the temporal evolution of a stellar model of a given mass.} in the mass range [0.4, 1.5] M$_{\odot}$ (in steps of 0.1 M$_{\odot}$) from the pre-main sequence phase up to the sub-giant branch and isochrones\footnote{The isochrone is the theoretical counterpart of an observed stellar cluster in the
Hertzsprung-Russell (\hr)/ Colour-Magnitude (\texttt{CM}) diagram, i.e. the locus of evolutionary models with different mass but with the same age and chemical composition; it is computed from the evolutionary tracks.} for ages from 1 to 12 Gyr, to cover all the possible cluster ages of interest for this analysis. For completeness, the differential analysis of the effects of the p-p rate variation has been performed for two different chemical compositions  ([Fe/H] = $+0.0$ and [Fe/H] = $-1.0$ as discussed in Sec.~\ref{sec:models}). 

Low-mass stellar tracks (or isochrones of old clusters) present at the central H-exhaustion the \tof{} feature, which is theoretically identified as the track/isochrone (Main Sequence) point with the highest effective temperature. However the \tof -region in the \hr/\texttt{CM} diagrams is almost vertical at the \tof -point (i.e., there is a large luminosity variation at essentially the same effective temperature) and this makes the precise identification of the \tof{} luminosity quite difficult. Thus, to reduce the intrinsic luminosity variation of the canonical \tof{}, following a technique similar to the one adopted in other works \cite{buonanno98,chab96,valle13a,valle13b}, we decided to measure the luminosity of a point brighter and with an effective temperature of 100 K lower than the \tof. 
Stars with intermediate and high mass (or isochrones with intermediate/low ages) show at the H-exhaustion the \oc{} feature. We defined the \oc{} similarly to the \tof{} point.

\begin{figure}[!t]
   \centering
 	\includegraphics[width=0.9\columnwidth]{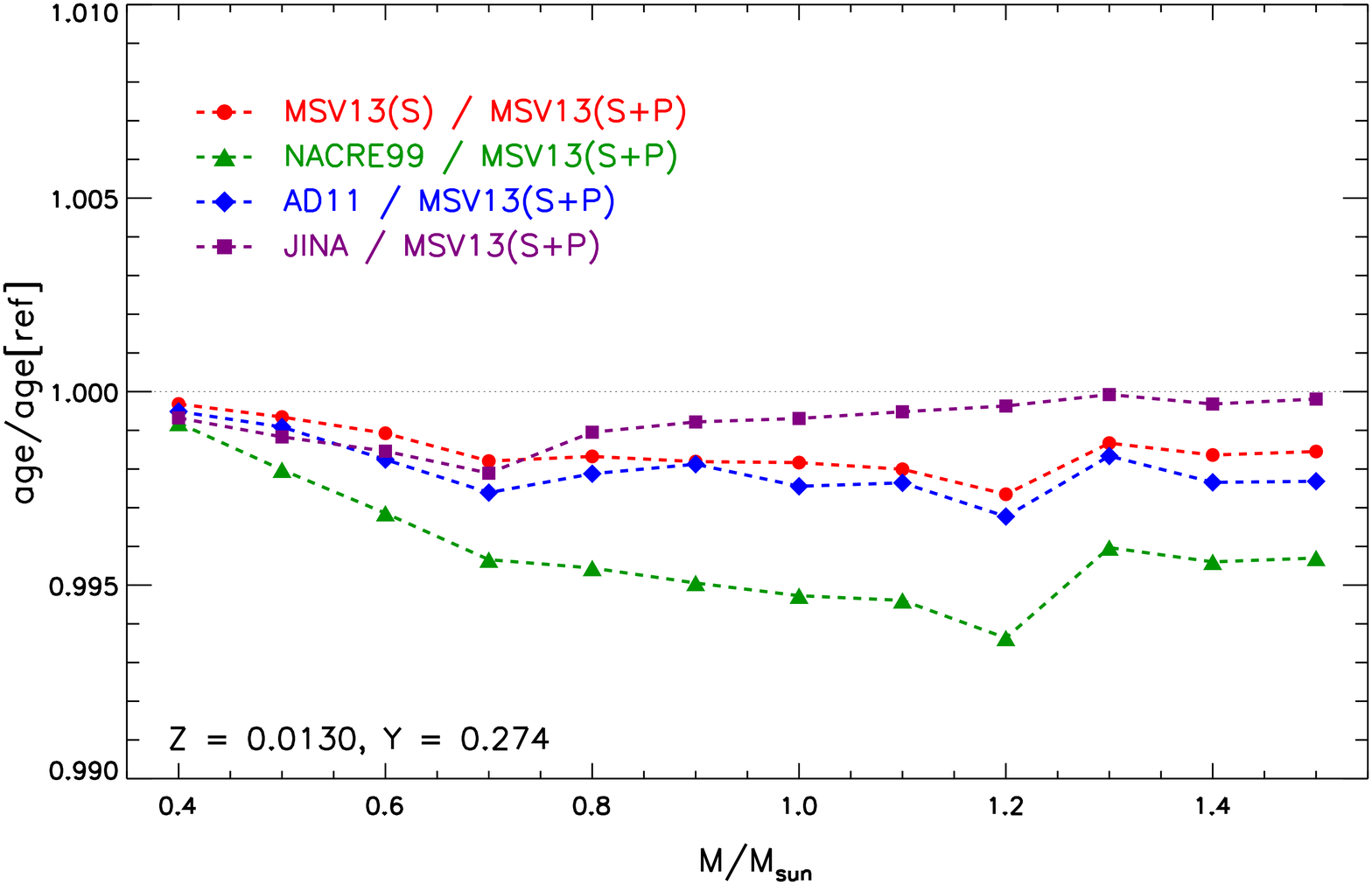} 
 	\includegraphics[width=0.9\columnwidth]{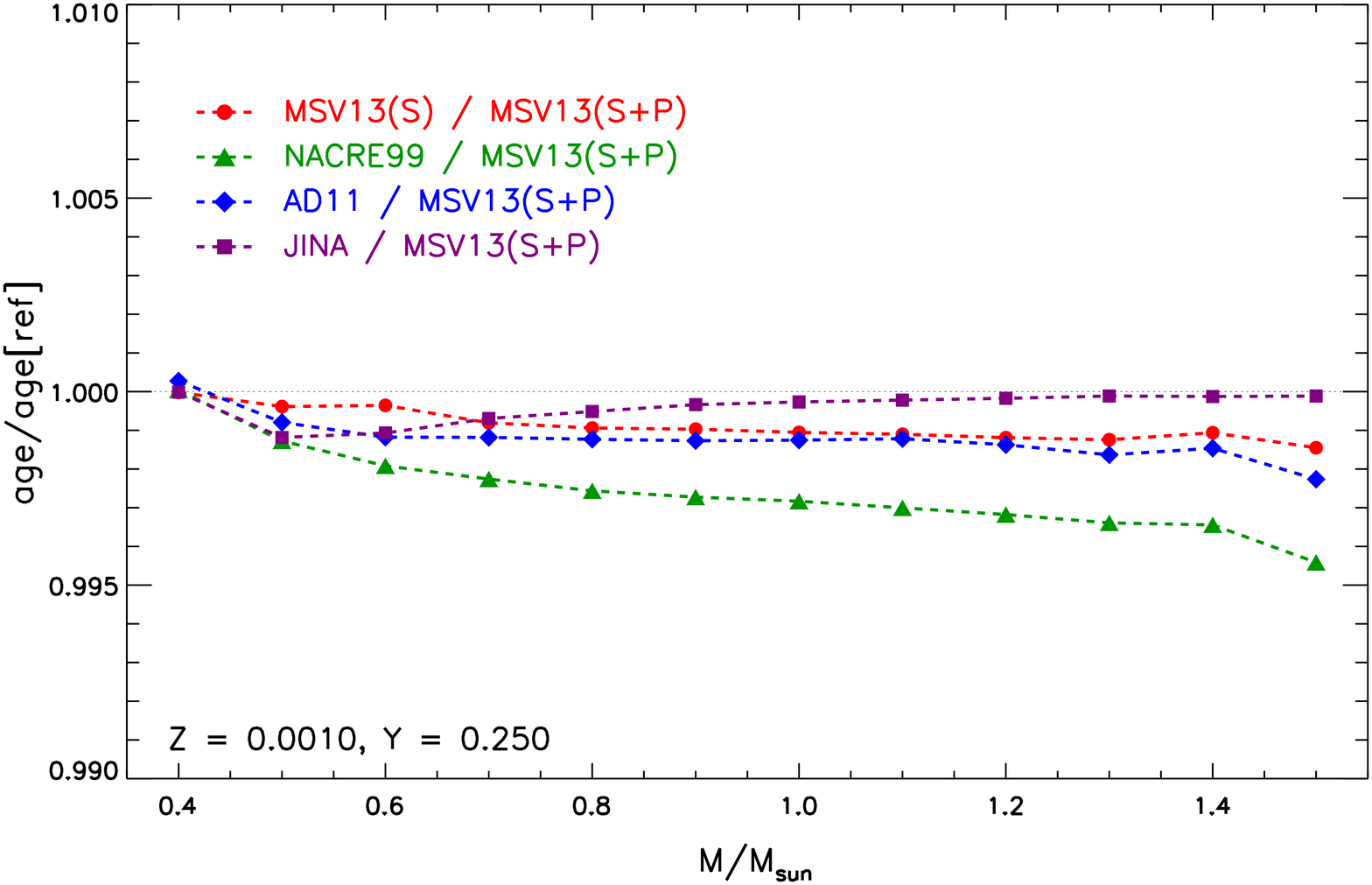}
  	\caption{Ratio of the H-exhaustion time, as a function of the stellar  mass, calculated for the labelled p-p rates to the one calculated with the reference reaction rate, \msv\texttt{(S+P)}. Upper panel: $Y=0.274$ and $Z=0.013$ (solar values); bottom panel: the same but for $Y=0.250$ and $Z=0.001$.}
    \label{ageTO}
\end{figure}

Figure \ref{ageTO} shows the ratio of the H-exhaustion time calculated for the rates analysed in Sec.~\ref{sec:rates} to the one calculated with the reference p-p rate, as a function of the stellar mass, for the labelled chemical compositions. As expected, the \nacre{} rate, which is the most different with respect to the reference one, leads to the maximum difference in the H-exhaustion time; however, even in this case the relative change is lower than about $5 \div 6 \permil$. Neglecting the contribution of the P-partial wave in the rate calculation gives an effect lower than about $2 \permil$.

The same quantities have been calculated for [Fe/H]=$-1.0$ ($Y=0.250$ and $Z=0.001$, bottom panel of Fig. \ref{ageTO}). In this case, the sensitivity to the p-p rate is slightly reduced, in agreement with the results obtained by Valle et al. \cite{valle13b}, which showed that the evolutionary effects of a variation of the p-p rate are very mildly affected by reasonable changes in helium and metal abundances.
Differences in the track \tof{}/\oc{} luminosity are always lower than $1 \permil$ and thus completely negligible. 

%

\begin{figure}[!t]
   \centering
 	\includegraphics[width=0.9\columnwidth]{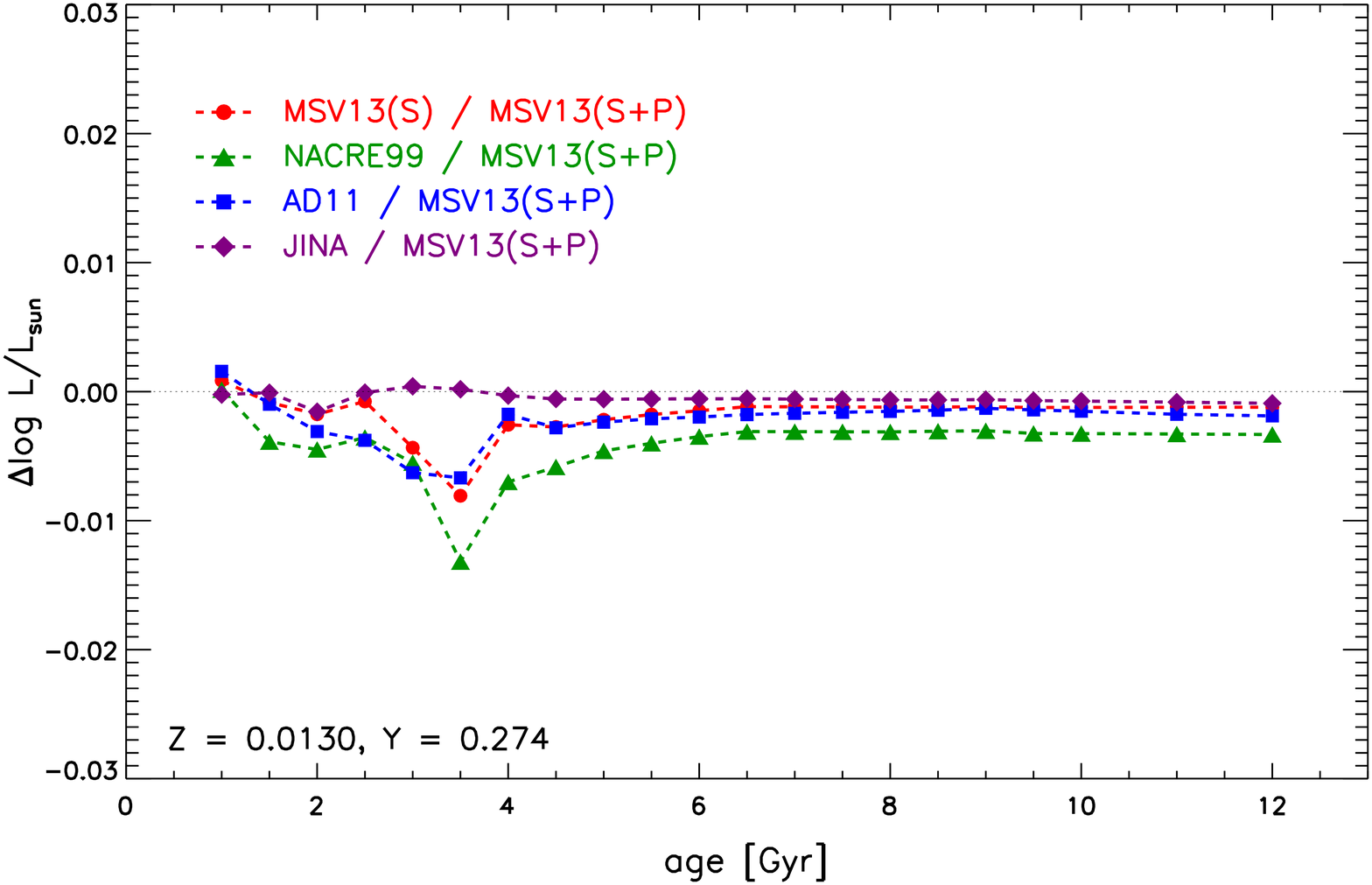}
  	\includegraphics[width=0.9\columnwidth]{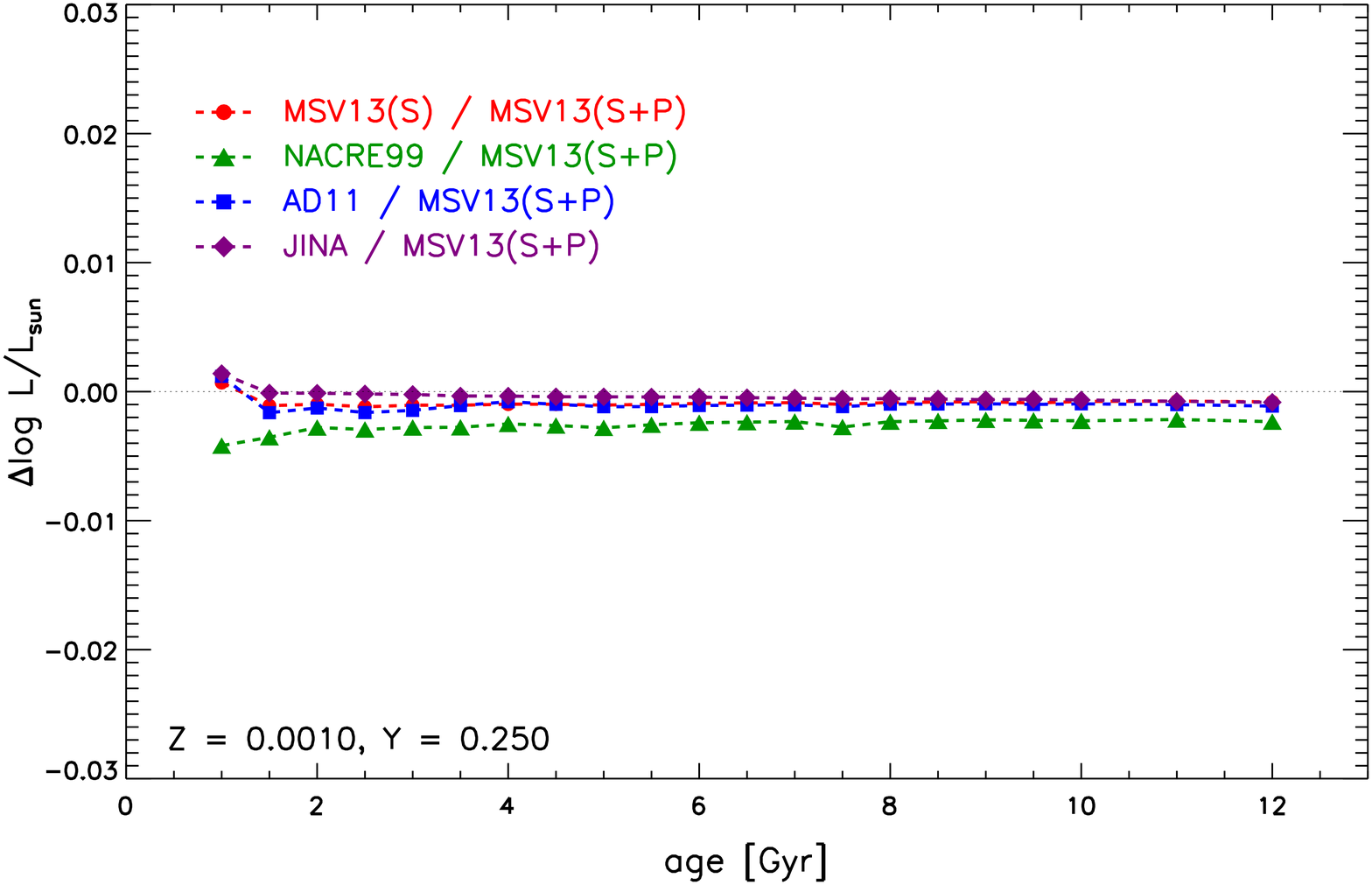} 
  	\caption{Luminosity difference at the \tof /\oc{} point, as a function of the cluster age, for isochrones calculated with the labelled p-p rates and the ones calculated with the \msv\texttt{(S+P)} one. Upper panel: $Y=0.274$ and $Z=0.013$ (solar values); bottom panel: the same but for $Y=0.250$ and $Z=0.001$.} 
   \label{TOisochrone}
    \end{figure}

Figure \ref{TOisochrone} shows the luminosity differences at the \tof /\oc{} for the isochrones as a function of the age, between models computed with the quoted p-p reaction rates and the reference one. For the solar chemical composition (upper panel), the differences in the \tof /\oc{} luminosity are lower than about $5 \permil$ for the most of cluster ages, reaching a value of 1-1.5$\%$ only for ages in the range 2-4 Gyr.  The major sensitivity to a p-p rate change for clusters in the age range 2-4 Gyr is due to the fact that the H-exhaustion region of the \hr{} diagram of these clusters is populated by stars of masses in the range [1.1, 1.5] M$_{\odot}$. These are just the transition masses (the precise value depending also on the stellar chemical composition) among stars which burn hydrogen in the central regions mainly through the p-p chain (lower main sequence stars) or through the CNO-cycle (upper main sequence stars). Such a different H-burning produces also a different track/isochrone morphology close to the \tof /\oc{} region. Thus, it is not surprising that the effect of the p-p rate change in the \tof /\oc{} luminosity is larger in this age range. In any case, we want to emphasise that such an effect is very small and it has a minor relevance in cluster age determination when compared to the other sources of uncertainty (see e.g. Refs.~\cite{chab96,chab98,krauss03,deglinnocenti09,valle13a,valle13b}).

The bottom panel of Fig. \ref{TOisochrone} shows the same quantities as the top one but for $Y=0.250$ and $Z=0.001$. In agreement with the results found for the stellar tracks, the sensitivity to a change of the p-p rates is slightly reduced and the feature in the region 2-4 Gyr is no more visible.

Given the above results and the very small uncertainty on the \msv{} $S(E)$, we can conclude that the p(p,e$^+ \nu_e$)$^2$H reaction rate is now known with such a high precision that it does not constitute any more a significant uncertainty source in the age evaluation of stellar clusters.

\section{The Sun and the  p(p,e$^+ \nu_e$)$^2$H cross section}
\label{sec:sun}

The Sun is unique among stars because several observational quantities are known with a very high precision. Thus, it is clear that the comparison between a theoretical solar model (Standard Solar Model, \ssm) and the actual Sun is a strong test of the validity of theoretical stellar computations. 

An \ssm{} is defined as a 1 M$_\odot$ model which reproduces, at the age of the Sun (t$_\odot$), within a given numerical tolerance, the observed properties of the Sun, by adopting a set of input physics (see e.g. Refs.~\cite{bahcall89, stix89} and references therein).

The present \ssm{} has been computed adopting M$_\odot = 1.989\times10^{33}$ g, L$_\odot=3.8418\times10^{33}$ erg s$^{-1}$, and R$_\odot = 6.9598\times 10^{10}$ cm \cite{serenelli11}. Regarding the age of the Sun, we have adopted t$_\odot$ = $(4.566 \pm 0.005)$ Gyr, as estimated from age determination for meteorites combined with models of the solar system formation \cite{bahcall95}. We have used the recent spectroscopically determined ratio of metals-to-hydrogen in the solar photosphere by Asplund et al. (2009) \cite{asplund09}, which results in $(Z/X)_{ph,\odot}= 0.0181$. We emphasise that the present observed photospheric composition is different from the initial one due to microscopic diffusion, whose efficiency must be theoretically estimated. Moreover, present photospheric helium abundance is not strongly constrained by direct observations, since helium lines are not observed in photosphere, and the external convection efficiency cannot be theoretically obtained in a firm way. Thus, one has the freedom of adjusting the initial helium, $Y$, metallicity, $Z$, and external convection efficiency required in the calculation. 

We have evolved an initially homogeneous solar mass from the pre-main sequence phase up to the solar age. To obtain L$_\odot$, R$_\odot$ and $(Z/X)_{ph,\odot}$ at t$_\odot$, we have tuned, by means of an iterative procedure, three parameters: the initial helium abundance $Y$ (mainly influencing the luminosity), the initial metal abundance $Z$ (mainly affecting the present $(Z/X)_{ph,\odot}$ surface value) and the $\alpha_\mathrm{ML}$ parameter of the mixing length theory \cite{bohm58}, related to the external convection efficiency (mainly influencing radius/effective temperature). The precision with which luminosity, radius and present $(Z/X)_{ph,\odot}$ surface value are reproduced in our \ssm{} is better than, respectively, $10^{-5}$, $10^{-4}$, and $4\times 10^{-4}$. 
\begin{table*}[!ht]
	\centering
	\begin{tabular}{l|c||rrrr}
 	 & \msv\texttt{(S+P)} & \msv\texttt{(S)} & \nacre & \ad & \jina \\
 	 & \multicolumn{1}{c||}{reference} & \multicolumn{4}{c}{relative differences}\\
 	\hline
	\hline
	T$_\mathrm{c}\,[10^7\,\mathrm{K}]$ & $ 1.54794$ & $-1\permil$ & $-3\permil$ & $ -2\permil$ & $-1\permil$ \\
	\hline
	$\Phi_\mathrm{pp}^\nu\,[10^{10}]$ & $ 6.020$ & $ 1\permil$ & $ 2\permil$ & $ 2\permil$ & $ 1\permil$ \\
	$\Phi_\mathrm{pep}^\nu\,[10^{8}]$ & $ 1.446$ & $-2\permil$ & $-6\permil$ & $ -2\permil$ & $-1\permil$ \\
	$\Phi_\mathrm{hep}^\nu\,[10^{3}]$ & $ 8.584$ & $-1\permil$ & $-3\permil$ & $ <1\permil$ & $ 2\permil$ \\
	$\Phi_\mathrm{Be-7}^\nu\,[10^{9}]$ & $ 4.503$ & \textbf{$-$1\%} & \textbf{$-$3\%} & \textbf{$-$1\%} & $-9\permil$ \\
	$\Phi_\mathrm{B-8}^\nu\,[10^{6}]$ & $ 3.694$ & \textbf{$-$3\%} & \textbf{$-$7\%} & \textbf{$-$4\%} & \textbf{$-$2\%} \\
	$\Phi_\mathrm{N-13}^\nu\,[10^{8}]$ & $ 2.417$ & \textbf{$-$2\%} & \textbf{$-$6\%} & \textbf{$-$3\%} & \textbf{$-$1\%} \\
	$\Phi_\mathrm{O-15}^\nu\,[10^{8}]$ & $ 1.811$ & \textbf{$-$3\%} & \textbf{$-$8\%} & \textbf{$-$4\%} & \textbf{$-$2\%} \\
	$\Phi_\mathrm{F-17}^\nu\,[10^{6}]$ & $ 3.373$ & \textbf{$-$3\%} & \textbf{$-$8\%} & \textbf{$-$4\%} & \textbf{$-$2\%} \\
	\hline
	\hline
	\end{tabular}
	\caption{The first column lists the results for central temperature (K) and neutrino fluxes (s$^{-1}$\,cm$^{-2}$) of our reference model (i.e. calculated with the \msv\texttt{(S+P)} p-p rate). The others columns list the relative difference between the results obtained for the reference \ssm{} and for \ssm{}s computed with the labelled p-p rates. Bold font: relative differences above $1\%$.}
\label{tab:sun}
\end{table*}

The observed solar luminosity fixes the efficiency of the p-p chain, whose energy production must counterbalance the energy losses from the surface (with a contribution of about $1\%$ from the CNO-cycle). Thus, the temperature (and the density) in the solar interior, needed to produce the required energy, depends on the p(p,e$^+ \nu_e$)$^2$H cross section. An increase of the p(p,e$^+ \nu_e$)$^2$H cross section would lead to a decrease of the stellar temperature, thus directly affecting the neutrino fluxes 
(see e.g. Refs.~\cite{bahulr88,bahcall89,castellani93,castellani97,ricciber97,deglinnocenti98,schlattl99,antia99,serenelli13}). The p-p neutrinos (together with $pep$ and $hep$ neutrinos) are directly connected to the solar luminosity and thus they are expected to be very weakly dependent on p(p,e$^+ \nu_e$)$^2$H cross section change. On the other hand, all the other neutrino fluxes are much more sensitive to temperature variations and, consequently, are more affected by the adopted p-p reaction rate (see e.g. Ref.~\cite{bahcall89}).
 
We calculated \ssm{}s by adopting as p(p,e$^+\nu_e$)$^2$H reaction rate the different evaluations discussed in Sec.~\ref{sec:rates}, keeping fixed all the other input physics. As reference model we have adopted the one calculated with the \msv\texttt{(S+P)} p-p rate. Among all the models, the differences in the original helium and metallicity abundances and in the mixing length value required to obtain a standard solar model are negligible.

As expected, being the solar luminosity fixed by observations, the solar central temperature decreases when the p(p,e$^+ \nu_e$)$^2$H reaction efficiency increases to counterbalance the increased nuclear energy production (see e.g. Ref.\cite{bahulr88}). The largest difference in the central temperature (for the \ssm{} with p-p rate from the \nacre{} compilation) is of the order of $3 \permil$. Due to the very small differences in the chemical composition and central temperature, all the calculated models are not expected to show variations in the predicted helioseismic quantities; we checked that this is the case at the level of few $\permil$. However, due to the high temperature dependence of neutrino fluxes but the p-p one, the effect on solar neutrinos is not totally negligible, even if small. 

Table \ref{tab:sun} shows the relative differences for the solar central temperature and the solar neutrino fluxes. The maximum difference for the solar neutrino fluxes corresponds to the adoption of the \nacre{} p-p reaction rate, reaching a maximum of about 8$\%$. This is not a negligible difference, however, as discussed in Sec.~\ref{sec:rates}, in our opinion the adoption of the NACRE99 p-p rate should be discouraged. The $\sim 1\%$ effect of the P-partial waves to the p-p reaction rate turns out into a maximum of $3\%$ effect on the $^8$B, $^{15}$O and $^{17}$F neutrino fluxes. 

As for the stellar cluster age, the very small error on the present p(p,e$^+ \nu_e$)$^2$H rate determination (see Sec.~\ref{sec:rates}) has a negligible effect on \ssm{} calculations.

%

\section{Summary}\label{sec:conclusion}

An updated p(p,e$^+ \nu_e$)$^2$H reaction rate has been calculated by adopting the evaluation of the astrophysical $S$-factor by Marcucci et al. \cite{marcucci13}, which takes into account two-photon and vacuum polarisation contributions and, for the first time, it includes all the P-partial waves in the incoming p-p channel. The uncertainty on the rate evaluation is now estimated at the level of few $\permil$. A release is available (at the link: \url{http://astro.df.unipi.it/stellar-models/pprate/}) for the calculation of the present p-p rate (\msv) with the possibility to select the energy range and resolution. 

The comparison with other widely adopted p-p rates shows maximum differences of about 3-4$\%$. The effects on stellar cluster age determination of the adoption of different p(p,e$^+ \nu_e$)$^2$H reaction rates have been discussed. We found that the maximum variation is obtained for the still widely used p-p rate by the \nacre{} compilation; however, taking into account the other uncertainty sources, this difference is of minor relevance for cluster age determination.

The influence of the adoption of different p-p rate prescriptions on the standard solar models calculations has been also analysed. The change of the p-p rate evaluation is negligible for the solar structure characteristics. However, the related tiny changes in the central temperature reflect in small, but not negligible, variations of the neutrino fluxes, but the p-p one, due to their high sensitivity to temperature variations.

Finally, we have also shown that the p(p,e$^+ \nu_e$)$^2$H reaction rate is now obtained with such a high accuracy that it does not constitute anymore a significant uncertainty source in stellar models.

\section*{Acknowledgement}
We warmly thank the anonymous referee for a careful reading of the manuscript and useful suggestions which improved the paper. This work has been supported by PRIN-MIUR 2010-2011 ({\em Chemical and dynamical evolution of the Milky Way and Local Group galaxies}, PI F. Matteucci), PRIN-INAF 2011 ({\em Tracing the formation and evolution of the Galactic Halo with VST}, PI M. Marconi), and  PRIN-INAF 2012 ({\em The M4 Core Project with Hubble Space Telescope}, PI L. Bedin). 

%
\bibliographystyle{elsarticle-num-usr}
\bibliography{bibliography}

\end{document}